\documentclass[11pt,a4paper,reqno]{amsart}
\usepackage{tikz}
\usepackage{verbatim}
\usepackage{amsmath}
\usepackage[all,cmtip]{xy}
\usepackage{tikz-cd}
\usepackage{multirow}
\usepackage[centertags]{amsmath}
\usepackage{amsfonts}
\usepackage{amssymb}
\usepackage{amsthm}
\usepackage{newlfont}
\usepackage{mathtools}
\usepackage{a4}
\usepackage{enumerate}
\usepackage{cite} 
\usepackage{diagbox}

\newcommand{\ben}{\begin{equation}}
\newcommand{\een}{\end{equation}}
\newcommand{\bens}{\begin{equation*}}
\newcommand{\eens}{\end{equation*}}
\newcommand{\nn}{\nonumber\\ }

\newcommand{\qq}{\qquad\qquad}

\newcommand{\pq}{\partial_q}
\newcommand{\pp}{\partial_p}
\newcommand{\Lpq}{\stackrel{\leftarrow}{\partial_q}}
\newcommand{\Lpp}{\stackrel{\leftarrow}{\partial_p}}
\newcommand{\Rpq}{\stackrel{\rightarrow}{\partial_q}}
\newcommand{\Rpp}{\stackrel{\rightarrow}{\partial_p}}
\newcommand{\Lp}{\stackrel{\leftarrow}{\partial}}
\newcommand{\Rp}{\stackrel{\rightarrow}{\partial}}
\newcommand{\cA}{{\cal A}}\newcommand{\cW}{{\cal W}}
\newcommand{\cD}{{\cal D}}

\newcommand{\I}{{\cal I}}\newcommand{\cQ}{{\cal Q}}
\newcommand{\cP}{{\cal P}}\newcommand{\cM}{{\cal M}}
\newcommand{\cO}{{\cal O}}

\newcommand{\hfb}{{\hfill\break}}
\newcommand{\sub}[1]{_{\hskip-1pt\stackrel{}{#1}}}

 \def\R{{\mathbb R}}  \def\N{{\mathbb N}}
   
\def\I{{\mathbb I}}

\usetikzlibrary{matrix}
\begin{document}
\baselineskip=18pt

\title[Augmented Star Products]{Can Star Products be Augmented by Classical Physics?}

\author[M.P.G. Robbins]{Matthew P.G.  Robbins$^{1,2}$}
\author[M.A. Walton]{Mark A. Walton$^2$}

\date{\today}
\vskip1cm
\begin{abstract}
\large\baselineskip=16pt 
${}$ \\ It has been suggested that star products in phase-space quantization may be  augmented to describe additional, classical effects. That proposal is examined critically here. Two known star products that  introduce classical effects are: the generalized Husimi product of coarse-grained quantization, and a non-Hermitian damped star product for the harmonic oscillator. Following these examples, we consider products related by transition differential operators to the classic Moyal star product. We restrict to Hermitian star products, avoiding problems already pointed out for the original damped product. It is shown, however, that with such star products, augmented quantization is impossible, since an appropriate classical limit does not result.  

For a more complete study, we then also consider generalized, or local, transition operators, that depend on the local phase-space coordinates, as well as their derivatives. In this framework, one example of possible physical interest is constructed. Because of its limited validity and complicated form, however, it cannot be concluded that augmented quantization with local transition operators is practical.  \\ 
\end{abstract}

\maketitle
\noindent
$^1$
Department of Physics and Astronomy, University of Waterloo, Waterloo ON, Canada N2L 3G1;\\
${}\ \,$ 
Perimeter Institute for Theoretical Physics, Waterloo ON, Canada, N2L 2Y5 $^3$\\
$^2$ 
Department of Physics and Astronomy, University of Lethbridge, Lethbridge AB, Canada T1K 3M4\\

\noindent
\textit{E-mail:} m2robbins@uwaterloo.ca, walton@uleth.ca
\hfb $^3$  Current addresses.
\vfill\eject

\section{Introduction} 

Observables in quantum mechanics can be described by operators or by functions (and distributions) in phase space \cite{Zachos2005, Cohen2013, Hillery1984, Balazs1983, Lee1995, Bayen1978a, Bayen1978b, Osborn1995,Takahashi1989}.  Operator and phase-space quantum mechanics are equivalent, however, as can be demonstrated through maps from one to the other. A  quantization map takes the phase-space quantum mechanics to the operator version, and its inverse is known as the dequantization map. The non-commutativity of operator observables is reflected in phase space by the non-commutative star product that must be used to multiply phase-space obervables. 

Quantization of a classical system is not unique either way, however.  Different quantizations give rise to different star products \cite{Bayen1978a, Bayen1978b, Cohen2013, Lee1995, Mehta1964}.  For example, distinct operator-ordering rules determine distinct quantization maps.  The corresponding dequantizations prescribe distinct star products.  Weyl operator ordering and the Moyal star product are paired, but so are standard operator ordering and the standard star product,  Born-Jordan ordering and another star product, etc.  

An advantage of phase space quantization (also known as deformation quantization) is that different quantizations and corresponding star products may be related in a simple way. For example, a transition (differential) operator \cite{Bayen1978a, Bayen1978b} can associate the phase-space observables and star products of 2 quantizations. 

The same mathematical machinery can introduce physical effects distinct from quantization, however.  Coarse graining in phase space is described by the generalized Husimi star product \cite{Husimi1940, Takahashi1989}.  It is obtainable from the Moyal product, e.g., by a transition differential operator. Similarly, another transition operator produces a star product that converts the equation of motion of the harmonic oscillator into that of the damped harmonic oscillator \cite{Dito2006, Belchev2009}.  

Here we study such star products modified to include additional classical effects not described by the original Hamiltonian. This scheme was suggested originally in \cite{Dito2006}. The damped star product introduced there was shown in \cite{Belchev2009} to have serious problems, however, because it is non-Hermitian. But we will see here that there exists a Hermitian damped star product similar to the original non-Hermitian one described in \cite{Dito2006}. We also note that the (generalized) Husimi star product introduces a classical coarse-graining effect, and is described by a transition differential operator in a similar way. It seems then, that the scheme might be viable. 

For brevity, the construction of such augmented star products will be called augmented (phase-space) quantization.  Taking the coarse-grained and damped star products as our guides, we undertake a serious study of augmented star products related to the Moyal star product by transition operators. Our goal is to see whether or not such augmented star products can provide effective descriptions of certain quantum systems. 

We find that augmented, Hermitian star products described by transition differential operators cannot. The classical limit is problematic. Only that part of the transition differential operator that is independent of Planck's constant is relevant to the classical limit. If it is 1, the  classical limit is not modified -  there is no additional  physics (no augmentation).  If the $\hbar$-independent part is non-trivial, then a multiplicative, rather than additive, modification of the equations of motion results. 

Augmented quantization is therefore not possible unless the framework is changed. For a more thorough treatment, we therefore go on to consider  generalized transition operators, depending not only on derivatives of phase-space coordinates, but also on the coordinates themselves.  This generalization is quite drastic - although they have been considered before (see \cite{Crehan1989, Pinzul2008}), such ``local'' transition operators are rarely invoked.  In this framework, one example of possible physical interest is constructed: a Hermitian phase-space star product for the weak damping of a simple harmonic oscillator.  This result is of limited validity and has a rather complicated form, however.  The practical feasibility of using local transition operators in augmented quantization remains to be demonstrated. 

Star products have been and continue to be very useful tools in theoretical physics.  Before we outline the organization of our paper, let us provide recent points of contact with the vast literature on star products in deformation quantization and other subjects. Reference \cite{Lizzi2014} includes a recent review in the context of this paper, non-relativistic quantum mechanics. A formal star product and associated quantization has been described for any finite-dimensional Poisson manifold as phase space in \cite{Kontsevich2003}; a generalization \cite{Dherin2015} has more recently been described for when the Lie algebra of classical observables is replaced by a Leibniz algebra.  Quantum field theory \cite{Dito1989} invokes star products with ordinary derivatives replaced by functional derivatives. Star products can encode the non-commutativity of non-commutative geometry \cite{Szabo2003}.  String theory in certain limits is described by such non-commutative geometry \cite{Seiberg1999} (see \cite{Blumenhagen2014} for a recent, pedagogical survey).  In a similar way, associativity is lost in certain regimes of string theory, motivating the very recent study of non-associative star products \cite{Kupriyanov2018, Szabo2018}.

The next section of this paper is a quick introduction to the elements of phase-space quantization that are relevant to our study, and to our notation. Section 3 describes the 2 star products we take as our guides: the coarse-grained Husimi \cite{Husimi1940, Takahashi1989} and damped \cite{Dito2006, Belchev2009} products. 

Section 4 treats augmented star products abstractly and generally. In subsection 4.1 we find that transition differential operators (``global'' transition operators) do not lead to classical equations of motion augmented by additional terms; this is our main result.  Subsection 4.2 demonstrates that local transition operators can, in principle, furnish new examples of modified star products that incorporate additional classical physics.  A single, simple example is constructed: a new, Hermitian  damped star product. It is, however, rather unwieldy, and we therefore believe that local transition operators may have limited usefulness. 

The final section is our conclusion.  

\section{Phase-space quantum mechanics and star products} 

We will restrict to 1-dimensional systems on position space $\R$ with coordinate $q$, and conjugate momentum $p\in \R$, so that the phase-space is $\R^2$. It is straightforward to generalize to several degrees of freedom.  Only time-independent Hamiltonians will be treated:  $\partial\sub{\,t}H=0$. 

A brief account of quantization in phase-space will now be given.  Subsection \ref{subsec: Moyal} then outlines the canonical example, involving the Weyl map, the Moyal star product, and the Wigner transform.  It will be our reference quantization, so that all other examples treated here will be related to it by transition operators, which we discuss in Subsection \ref{subsec: transition operators}. 

\subsection{Quantization in Phase-Space}

Suppose that the distribution on phase-space, $f=f(q,p)$, is a classical observable.  Then the quantization map $\cQ$ produces a quantum observable:
\begin{align} 
\cQ\big( f(q,p) \big)\ =\ \hat f\ \ .\label{Qn}
\end{align}
The $\hspace{0.05cm}\hat{}\hspace{0.05cm}$ indicates that $\hat f$ is an operator, which is a function of position operator, $\hat q$, and momentum operator, $\hat p$.  The inverse (dequantization map) $\cW$ is given by 
\begin{align}
\cW(\hat f )\ =\ f(q,p)\ .
\label{cW}\end{align}
Strictly, both $\cQ$ and $\cW$ should be labelled by the phase-space coordinates, so that 
\begin{align}
\cQ^{(q,p)}\Big( f(q,p)  \Big)\ =\ \hat f\ ,\ \ \ \ \cW\sub{(q,p)}\Big( \hat f \Big)\ =\ f(q,p)\ . 
\label{QWqp}\end{align}
To avoid overly cumbersome notation, however, we will follow convention and drop these labels when confusion is unlikely.

Dequantization maps operators into phase-space distributions (the symbols of the operators). Operator products become star products, homomorphically:
\begin{align}
\cW\big(\, \hat f\, \hat g   \,\big)\ =\ \cW(\hat f)\, *\, \cW(\hat g)\ .
\label{homostar}\end{align} 
The star product $*$ is a bi-differential operator expressible in terms of the left derivatives $\Lpq, \Lpp$, and right derivatives, $\Rpq, \Rpp$, defined by 
\begin{align}
f(q,p)\Lpq g(q,p)\ :=\ \I(1,2)\,\,\partial_{q_1}\, f(q_1,p_1)\, g(q_2,p_2)\ , 
\label{defLpq}\end{align}
and in a similar manner for the right derivatives. Here $\I(1,2)$ enacts the identifications ${q_1\, =\, q_2\, =q}$ and ${p_1\,=\, p_2\, =\, p\,}.$
In useful shorthand notation,  (\ref{defLpq}) is  
\begin{align}
f\Lpq g\ :=\ \I(1,2)\,\partial_{q}(1)\, \Big( f(1)\, g(2) \Big)\,.
\label{shorth}\end{align}

As an illustration of the homomorphism of (\ref{homostar}), consider the Heisenberg-Weyl group relation in operator quantum mechanics: 
\begin{align}
\exp\left[ i(\varphi \hat q + \xi \hat p )/\hbar \right]\exp\left[ i(\varphi' \hat q + \xi' \hat p )/\hbar \right]=\text{e}^{-\frac i{2\hbar} \big( \varphi\xi'-\xi\varphi' \big)}\,\exp\left\{ i[ (\varphi+\varphi')\hat q + (\xi+\xi')\hat p ]/\hbar\right\}\,,
\end{align} 
where $\varphi, \xi, \varphi', \xi'\in\R$, a consequence of the simple Baker-Campbell-Hausdorff formula. Application of the dequantization map $\cW$  yields
\begin{align}
\cW\big(\exp\left[ i(\varphi \hat q + \xi \hat p )/\hbar \right]\big)\, *\,& \cW\big(\exp\left[ i(\varphi' \hat q + \xi' \hat p )/\hbar \right]\big)\ \nn =\ &\text{e}^{-\frac i{2\hbar} \big( \varphi\xi'-\xi\varphi' \big)}\, \cW\Big(\exp\left\{ i[ (\varphi+\varphi')\hat q + (\xi+\xi')\hat p ]/\hbar\right\}  \Big)\ .
\label{starHW}\end{align} 
This demonstrates that a phase-space quantization produces a $*$-realization of the Heisenberg-Weyl group. 

Let $\cD$ denote an arbitrary bi-differential operator (such as the star product $*$, or a left- or right-derivative $\Lpq, \Lpp$ or $\Rpq, \Rpp$, for examples). Transpose exchanges left- and right-derivatives:
\begin{align}
\big( \Lp \big)^{\, t}\ =\ \Rp\ ,\ \ \big( \Rp\big)^{\, t}\ =\ \Lp\ ,
\end{align} 
so the transpose $\cD^{t}$ satisfies 
\begin{align}
f\,\cD^t\,g\ =\ g\,\cD\,f\ ,
\label{cDtranspose}\end{align}
for arbitrary phase-space distributions (i.e. observables) $f$ and $g$. Let $\overline{\cD}$ and $\overline f$ be the complex conjugates of bi-differential operator $\cD$ and phase-space distribution $f$. The adjoint, or Hermitian conjugate, $\cD^\dagger$ of $\cD$ is the complex-conjugate transpose: $\cD^\dagger = \overline{\cD^{t}}$, so that 
\begin{align}
\overline{f\,\cD\,g}\ =\ \bar g\,\cD^\dagger\,\bar f\ .
\label{cDadjoint}\end{align}
A bi-differential operator, such as a star product $*$, can be Hermitian, $\cD^\dagger = \cD$, or symmetric, $\cD^t = \cD$, or real, $\overline{\cD} = \cD$, or none of the above. 

Application of $\cW$ to any relation involving operator observables yields the phase-space counterpart. With this in mind, consider the equation of motion for a quantum observable $\hat f$ in the Heisenberg picture,
\begin{align}
i\hbar\, \frac{\text{d}\hat f}{\text{d}t}\ =\ [ \hat f, \hat H ]\ , 
\label{HeisEMf}
\end{align}
assuming $\hat H^\dagger = \hat H$ is the Hamiltonian.  

As a result of (\ref{homostar}), commutators of operators are mapped to $*$-commutators,
\begin{align}
\cW\big(\, [ \hat f, \hat g  ]\,\big)\ =\ [ \cW\hat f, \cW\hat g ]\sub{*}\ :=\ \cW\hat f\, *\, \cW\hat g\, -\, \cW\hat g\, *\, \cW\hat f\ . 
\label{homobrack}
\end{align}
Applying the dequantization map $\cW$ to (\ref{HeisEMf}) yields 
\begin{align}
\dot f\ =\ \frac{[ f, H]\sub{*}}{i\hbar}\ =\ \{ f, H \}\sub{*}\ ,
\label{HeisMoyalEqn}
\end{align}
where $H:=\cW(\hat H)$, and we have introduced the Moyal bracket $\{\cdot, \cdot\}\sub{*}$:  
\begin{align}
\frac{[ f, g]\sub{*}}{i\hbar}\ =:\ \{ f, g \}\sub{*}\ =:\ f\, \cM\, g\ =\ f\left(\frac{*\, -\, *^{\, t}}{i\hbar}\right)g\ .
\label{MoyalB}\end{align}
$\cM$ denotes the associated Moyal bi-differential operator. 

The formal solution to (\ref{HeisMoyalEqn}) is 
\begin{align}
f(q,p;t)\
 =\ f(q,p;0)\, \exp\{\cM H t\}\ =\ \text{e}^{\frac{it}{\hbar} H*}\, f(q,p;0)\, \text{e}^{- * \frac{it}{\hbar} H}\ =\ U(-t) * f(q,p;0) * U(t)\ , 
\label{intHMEqn}
\end{align}
where  
\begin{align}
U(t)\
 =\ \exp\sub{*}\Big\{- \frac{it}{\hbar} H \Big\}\ =\ \cW\big( \text{e}^{-it\hat H/\hbar} \big)\ 
\label{propU}
\end{align}
is the symbol of the propagator, and $\exp\sub{*}$ indicates the $*$-exponential \cite{Bayen1978a,Bayen1978b}.

In the so-called classical limit $\hbar\to 0$, the Moyal bracket reverts to a Poisson bracket, 
\begin{align}
\lim_{\hbar\to 0}\, \{ f, g \}\sub{*}\ =\ \{ f, g \}\ . 
\label{MoyalBPoissonB}
\end{align}
The corresponding bi-differential operators have the same relation, 
\begin{align}
\lim_{\hbar\to 0}\, \cM\ =\ \cP\ .
\label{MoyalPDOPoisson}
\end{align}
Here the Poisson bi-differential operator $\cP$ is defined by the Poisson bracket:
\begin{align}
\{ f, g \}\ =:\ f\, \cP\, g\ ,\ {\text{i.e.}\,,}\ \ \cP\ =\ \Lpq\Rpp - \Lpp\Rpq\ .   
\label{PoissonPDO}
\end{align}
Therefore, in the classical limit, the equation of motion (\ref{HeisMoyalEqn}) reverts to 
\begin{align}
\dot f\ =\ \{ f, H \}\ ,
\label{EqMcl}\end{align}
which is the classical equation of motion, as expected.  

The quantum state of the system is described by the density operator in the Schroedinger picture, $\hat\rho$.  Its equation of motion is   
\begin{align}
i\hbar\, \frac{\text{d}\hat\rho}{\text{d}t}\ =\ [ \hat H, \hat\rho ]\ .  
\label{EMrho}
\end{align}
The dequantization map $\cW$ transforms the density operator $\hat\rho$ into a function $W(q,p;t)=\cW(\hat\rho)$ on phase space, obeying 
\begin{align}
i\hbar\, \frac{\partial W}{\partial t}\ =\ \{ H, W \}\sub{*}\ .  
\label{HeisEM}
\end{align}
$W(q,p;t)$ is called a quasi-probability distribution on phase space because, although it determines expectation values as a probability distribution would: 
\begin{align}
\langle A \rangle\ =\ \int {\text d}q\, {\text d}p\, W(x,p;t)\, A(x,p;t)\ ,
\label{expval}
\end{align}
it takes negative values. The first example of such a quasi-probability distribution is the Wigner function (see the next section).

Now that we have discussed the general form of phase-space quantization, we are in a position to identify the crux of our paper. We investigate the possibility that the star product may be augmented such that instead of (\ref{MoyalPDOPoisson}), we find 
\begin{align}
\lim_{\hbar\to 0}\, \cM\ =\ \cP\, +\, \delta\cP\ , 
\label{limMaugP}
\end{align}
where the additional term $\delta\cP$ describes extra, classical effects. We will return to this question in Section \ref{sec: Augmented star products}.  In the original example \cite{Dito2006}, the effect produced a damping term for the harmonic oscillator. 

We should point out that since 
\begin{align}
\cM\ =\ \frac{*\, -\, *^t}{i\hbar}\ ,
\label{cMstar}
\end{align}
(\ref{limMaugP}) is only possible because the pointwise multiplication of classical observables is modified in the classical limit:
\begin{align}
\lim_{\hbar\to 0}\, *\ \not=\ 1\ .
\label{limstarnotid}
\end{align}
Admittedly, this is a strange feature of augmented quantization. But it is  (\ref{limMaugP}) that determines the classical equations of motion, and physical examples are not ruled out by (\ref{limstarnotid}).

\subsection{Reference phase-space quantization}
\label{subsec: Moyal}

As our reference example, we will use the Weyl quantization map $\cQ_0$, which can be defined by 
\begin{align} 
\cQ_0\Big( \exp\left[ i(\varphi q + \xi p )/\hbar \right] \Big)\ =\  \exp\left[ i(\varphi \hat q + \xi \hat p )/\hbar \right]\ .  \label{Qzero}
\end{align} 
Expanding exponentials and equating terms proportional to $\varphi^n\, \xi^m$ produces the Weyl operator-ordering rule 
\begin{equation}
\begin{aligned}
\cQ_0\Big( q^n\, p^m \Big)\ &= \ \frac{1}{(n+m)!}\left(\hat{q}^n\hspace{0.03cm}\hat{p}^m+\hat{q}^{n-1}\hspace{0.03cm}\hat{p}^m\hspace{0.03cm}\hat{q}+\hat{p}\hspace{0.05cm}\hat{q}^n\hspace{0.03cm}\hat{p}^{m-1}+\hat{q}\hspace{0.05cm}\hat{p}\hspace{0.05cm}\hat{q}^{n-1}\hspace{0.03cm}\hat{p}^{m-1}\ + \ \cdots\right) \ , \\
&=\ \frac 1{(n+m)!} \sum_{\pi\in S_{n+m}} \stackrel{\pi(1)}{\hat q}\cdots\stackrel{\pi(n)}{\hat q}\,\,\stackrel{\pi(n+1)}{\hat p}\cdots\stackrel{\pi(n+m)}{\hat p}\ .
\label{Qzeronm}\end{aligned}
\end{equation}
Here the sum is over all permutations $\pi\in S_{n+m}$ of the $n+m$ factors in $\hat q^n \hat p^m$, and the numbers above the operators indicate their place in the product. The sum is somewhat redundant: it can be restricted to permutations $\pi$ in the coset $S_{n+m}/(S_n\times S_m)$, if $1/(n+m)!$ is replaced by $n!\, m!/(n+m)!$.  $\cQ_0(q^n\, p^m )$ is the average of distinct terms obtained by permuting the factors of $\hat q^n \hat p^m$.   Alternate expressions, such as  
\begin{align} 
\cQ_0\Big( q^n\, p^m \Big)\ =\ \frac 1 {2^n} \sum_{\ell=0}^m \binom{m}{\ell}\, \hat q^{\, n-\ell} \hat p^{\, m} \hat q^{\,\ell} =\ \frac 1 {2^m} \sum_{\ell=0}^n \binom{n}{\ell}\, \hat p^{\, m-\ell} \hat q^{\, n} \hat p^{\,\ell} \ ,
\end{align}
for example, can be derived using the Heisenberg commutation relation \cite{Cohen2013}. 

The dequantization map $\cW_0$, the inverse of the Weyl map (\ref{Qzero}),  
\begin{align} 
\cW_0\Big( \exp\left[ i(\varphi \hat q + \xi \hat p )/\hbar \right] \Big)\ =\  \exp\left[ i(\varphi q + \xi p )/\hbar \right]\ ,  \label{cWzero}
\end{align} 
is also known as the Wigner transform. The homomorphism (\ref{homostar}) between $*$- and operator products, along with the Heisenberg-Weyl relation (\ref{starHW}), gives the famous (Groenewold-)Moyal star product
\begin{align}
*\sub{\, 0}\ =\ \exp\left\{  \frac{i\hbar}{2} \left[ \Lpq\,\Rpp - \Lpp\,\Rpq \right] \right\}\ .
\label{starz}\end{align}
The transpose of the Moyal star product equals its complex conjugate, $\bar*\sub{\, 0}$:
\begin{align}
*\sub{\, 0}^{\, t}\ =\ \exp\left\{  \frac{i\hbar}{2} \left[ \Lpq\,\Rpp - \Lpp\,\Rpq \right] \right\}\ = \bar{*}\sub{\, 0}\ , 
\end{align}
so $*\sub{\, 0} = *\sub{\, 0}^{\,\,\dagger}$  is Hermitian. 

The Moyal product is associative: 
\begin{align}
\Big( f\, *\sub{\, 0}\, g \Big)\, *\sub{\, 0}\, h\ =\ f\, *\sub{\, 0}\, \Big( g \, *\sub{\, 0}\, h \Big)\ , 
\label{starzassoc}\end{align} 
which follows from 
\begin{align}
*\sub{\, 0}(1+2,3)\,*\sub{\, 0}(1,2)\ =\ *\sub{\, 0}(1,2+3)\,*\sub{\, 0}(2,3)\ .
\label{PREstarzassoc}\end{align}
Here the notation of (\ref{shorth}) is used, with $*\sub{\, 0}(1+2,3)$, for example,  obtained from $*\sub{\, 0}$ by replacing $\Lpq$ $\rightarrow$ $\partial_{q_1}+ \partial_{q_2}$, $\Lpp$ $\rightarrow$ $\partial_{p_1}+ \partial_{p_2}$, and $\Rpq, \Rpp,$  $\rightarrow$ $\partial_{q_3}, \partial_{p_3}$, respectively.

As the Moyal product (\ref{starz}) is
\begin{align}
*\sub{\, 0}\ =\ \exp\left\{ \frac{i\hbar}{2}\, \cP\, \right\}\ ,
\end{align}
the corresponding Moyal bracket bi-differential operator  (\ref{MoyalB}) is 
\begin{align}
\cM\sub{\, 0}\ =\ \frac{*\sub{\, 0}\ -\ *\sub{\, 0}^{\, t}}{i\hbar}\ =\ 2\, \sin\left(\hbar\, \cP\,/ 2 \right)/\hbar\ .
\label{cMzsin}\end{align}
In agreement with (\ref{MoyalPDOPoisson}), we find
\begin{align}
\lim_{\hbar\to 0}\, \cM\sub{\, 0}\ =\ \cP\ , 
\label{ClimPB}\end{align}
as expected. As already stated above, we will be interested in modifying this last relation to include additional classical physics.  

The reference quasi-probability distribution function is the famous Wigner function
\begin{align}
W\sub{0}(q,p;t)\
 =\  \frac{1}{2\pi}\int dy\,\text{e}^{-ipy}\bigg\langle q+\frac{\hbar y}{2}\Big|\,\hat{\rho}\,\Big|q-\frac{\hbar y}{2}\bigg\rangle\ .
\end{align}
Here $\hat{\rho}$ is the density operator, and $W\sub{0}(q,p;t) = \cW\sub{0}(\hat\rho)$.

\subsection{Other phase-space quantizations: transition differential operators}
\label{subsec: transition operators}

Phase-space quantization is not unique.  In many cases, however, the different quantizations can be related by a transition differential operator $T=T(\partial_q, \partial_p)$ \cite{Bayen1978a, Bayen1978b}. 

For example, an operator ordering different from the Weyl ordering of equations (\ref{Qzero}, \ref{Qzeronm}) may be used.  The Born-Jordan quantization map is 
\begin{align}
\cQ_{BJ}\Big( \exp\left[ i(\varphi q + \xi p )/\hbar \right] \Big)\ =\  \int_0^1d\alpha\, \text{e}^{i\alpha\varphi\hat q/\hbar}\, \text{e}^{ i\xi \hat p/\hbar}\, \text{e}^{i(1-\alpha)\varphi\hat q/\hbar}\ ,   
\label{QBJ}\end{align}
giving the ordering prescription \cite{Cohen2013}
\begin{align}
\cQ_{BJ}\Big( q^n\, p^m  \Big)\ =\  \frac 1{n+1}\, \sum_{k=0}^n \hat q^{\, n-k}\, {\hat p}^{\, m}\, \hat q^{\, k}\ .
\label{QBJnm}\end{align} 
The quantization maps $\cQ\sub{\, 0}$ and $\cQ_{BJ}$ are related. Applying the simple Baker-Campbell-Hausdorff formula to (\ref{QBJ}) yields 
\begin{align}
\cQ_{BJ}\Big( \exp\left[ i(\varphi q + \xi p )/\hbar \right] \Big)\ =\ \frac{\sin(\varphi\xi/2\hbar)}{\varphi\xi/2\hbar}\, \exp\left[ i(\varphi \hat q + \xi \hat p )/\hbar \right]\ ,
\end{align}
so that
\begin{align}
\cQ_{BJ}\Big( \exp\left[ i(\varphi q + \xi p )/\hbar \right] \Big)\ = \frac{\sin(\varphi\xi/2\hbar)}{\varphi\xi/2\hbar}\, \cQ\sub{\, 0}\Big( \exp\left[ i(\varphi q + \xi p )/\hbar \right] \Big)\ .
\end{align}

Both Weyl and Born-Jordan operator orderings are Hermitian.  A non-Hermitian example is the so-called standard operator ordering, with rule \cite{Lee1995}
\begin{align}
\cQ_{S}\Big( q^n\, p^m  \Big)\ =\   \hat q^{\, n}\, {\hat p}^{\, m}\ . 
\label{QSnm}\end{align} 
The relation to Weyl ordering is encoded in 
\begin{align}
\cQ_{S}\Big( \exp\left[ i(\varphi q + \xi p )/\hbar \right] \Big)\ =\ \exp\left( i\varphi \hat q/\hbar \right)\, \exp\left( i\xi \hat p/\hbar \right) \ =\ \text{e}^{\frac{-i\varphi\xi}{2\hbar}}\, \cQ\sub{\, 0}\Big( \exp\left[ i(\varphi q + \xi p )/\hbar \right] \Big)\ .
\label{QSQz}\end{align} 
Notice that this relation, for a non-Hermitian ordering, involves a complex multiplicative function. 

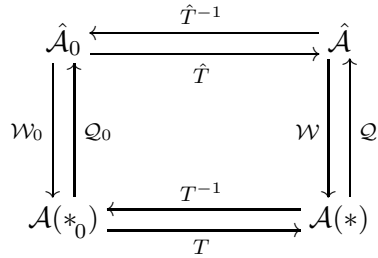
\begin{figure}[t!]
\begin{equation*}
\xymatrixcolsep{6pc}
\xymatrixrowsep{4pc}
\xymatrix{
   {\hat\cA_0} \ar@<-4pt>[r]_{\hat T} \ar@<-4pt>[d]_{{\cal W}_0} & {\hat\cA} \ar@<-4pt>[d]_{{\cal W}} \ar@<-4pt>[l]_{\hat T^{-1}}\\
   {\cA(*\sub{0})}\ar@<-4pt>[u]_{{\cal Q}_0} \ar@<-4pt>[r]_{T} & {\cA(*)} \ar@<-4pt>[l]_{T^{-1}} \ar@<-4pt>[u]_{{\cal Q}}  
}
\end{equation*}
\caption{For a fixed classical system,  consider a reference operator quantization map $\cQ_0$ and its related dequantization $\cW_0$. The algebra of quantum observables is the operator algebra $\hat\cA_0$ of operator quantum mechanics, or the $*\sub{0}$-algebra $\cA(*\sub{0})$ in phase-space. Suppose a different quantization map $\cQ$ and corresponding dequantization $\cW$ (with their accompanying algebras $\hat\cA$ and $\cA(*)$, respectively) are related to the reference quantizations through the transition operator $T$. The quantization and dequantization maps, and the  maps  between the observable algebras are shown. The arrows labelled by $\hat T$ and $\hat T^{-1}$ are included for completeness. $\hat T$ indicates what might be called the transition superoperator.  }
\label{figure:maps}
\end{figure}

Suppose $\cQ, \cW$ are the quantization and dequantization maps of a phase-space quantization. We can connect these maps to our reference quantization of $\cQ\sub{\, 0}, \cW\sub{\, 0}$ using an invertible differential operator $T$ such that  
\begin{align}
\cW\ =\ T\, \cW\sub{\, 0}\ ,\ \ \cQ\ =\ \cQ\sub{\, 0}\, T^{-1}\ .
\end{align}
We show pictorially how to relate different maps in Figure \ref{figure:maps}.

As both $\cW$ and $\cW\sub{\, 0}$ are homomorphisms from  operator products to $*$-products (see (\ref{homostar})), we find
\begin{align}
T\left(\, f\, *\sub{\, 0}\, g \,\right)\ =\ Tf\, *\sub{\, T}\, Tg\ ,  
\label{starT}\end{align} 
for any two phase-space distributions $f, g$.  This equation determines the product $*\sub{\, T}$. With $F=Tf$ and $G=Tg$, we have
\begin{align}
T\left(\, T^{-1}F\, *\sub{\, 0}\, T^{-1}G \,\right)\ =\ F\, *\sub{\, T}\, G\ .   
\label{starTFG}\end{align} 
Consider transition differential operators $T=T(\pq, \pp)$, i.e. those without any dependence on $q, p$. Since $F, G$ are arbitrary, (\ref{starT}) yields
\begin{align}
*\sub{\, T}(1,2)\ =\ T(1+2)\, *\sub{\, 0}\hskip-2pt(1,2)\,\, T^{-1}(1)\, T^{-1}(2)\ ,
\label{starTstar}\end{align}
using the notation introduced in (\ref{defLpq}), so that, because of the Liebniz rule,
\begin{align}
T(1+2)\ =\ T(\partial_{q_1}+\partial_{q_2}, \partial_{p_1}+\partial_{p_2})\ .
\end{align}
We therefore obtain 
\begin{align}
*\sub{\, T}\ =\ *\sub{\, 0}\,\, \odot\sub{\, T}\ ,
\label{starodot}\end{align}
with
\begin{align}
\odot\sub{\, T}\ =\ T^{-1}(\Lpq,\Lpp)\, T(\Lpq+\Rpq, \Lpp+\Rpp)\, T^{-1}(\Rpq, \Rpp)\; ,\  {\text{when}}\ T=T(\pq, \pp)\ .
\label{Tnoqp}\end{align}
This formula appeared in \cite{Belchev2009}. 

The Born-Jordan quantization discussed above provides an example:  the transition operator and star product 
\begin{align}
T\sub{BJ}\ =\ \text{sinc}\big(\,  \hbar\, \partial_p\partial_q /2 \,\big)\ \ ,\ \ \ *\sub{BJ}\ =\ *\sub{0}\  \, \frac{\, \text{sinc}\big[\, \hbar\, (\Lpp\, +\, \Rpp )\, (\Lpq\, +\, \Rpq) /2  \,\big] }{\text{sinc}( \hbar \Lpp\Lpq /2 )\, \text{sinc}( \hbar \Rpp\Rpq /2 ) }
\label{BJTstar}\end{align} 
are related by (\ref{starodot}, \ref{Tnoqp}). In this case,  we have a real transition operator, $T\sub{BJ} = \overline{T\sub{BJ}}$ and so a Hermitian star product $*\sub{BJ} = *\sub{BJ}^\dagger$. 

As an illustration of a non-Hermitian star product, consider the standard operator ordering of (\ref{QSnm}, \ref{QSQz}). Then the relevant transition operator and star product are 
\begin{align}
T\sub{S}\ =\ \exp\big[  i\hbar\, \partial_p \partial_q /2 \big]\ \ , \ \ \ *\sub{S}\ =\ \exp\big[ i\hbar\, \Lpq \Rpp \big]\ . 
\label{sTstar}\end{align}
Similarly, (\ref{starodot}, \ref{Tnoqp}) relate this non-real transition operator and non-Hermitian star product.

\section{Augmented star products: guiding examples}

Intriguingly, attempts have been made to use transition operators to introduce classical, physical effects, and not just to relate alternative quantizations.  We wish to examine the feasibility of this technique.

Two guiding examples are discussed in the following subsections. The first classical effect is Gaussian coarse-graining in phase-space, which gives rise to a (generalized) Husimi star product and phase-space quantization \cite{Husimi1940, Takahashi1989}. The second case  introduced damping into the simple harmonic oscillator equations of motion, and produced a modified star product depending on the damping coefficient \cite{Dito2006, Belchev2009}. 

\subsection{Coarse-grained Husimi quantization} 

Consider a distribution in phase-space, $f(q,p)$, coarse grained as follows:
\begin{align}
\frac 1{\pi\eta}\, \int dq'\, dp'\, f(q', p') \exp\left\{ - \frac 1 \eta\,\left[ \frac{(q-q')^2}{\sigma^2}  +  \sigma^2(p-p')^2  \right]  \right\}\, =\, \exp\left[ \frac \eta 4 \left( \sigma^2\pq^2 + \frac 1{\sigma^2} \pp^2  \right) \right] f(q,p)\ .
\label{coarse}\end{align}
Here $\eta$ is a classical coarse-graining scale, independent of $\hbar$, and $\sigma$ is a squeezing parameter. When $\eta=\hbar$, and $f$ is the Wigner function $W$, the expressions in (\ref{coarse}) equal the original Husimi quasi-probability distribution \cite{Husimi1940, Takahashi1989}. 

By (\ref{coarse}),
\begin{align}
T\sub{\,\eta}\ :=\ \exp\left[ \frac \eta 4 \left( \sigma^2\pq^2 + \frac 1{\sigma^2} \pp^2  \right) \right]
\label{Teta}\end{align}
can be interpreted as a transition differential operator. As (\ref{coarse}) is a coarse-grained Wigner function, it would be improper to say that the Husimi distribution is the result of an alternative quantization.  An additional classical physical effect (coarse-graining) is introduced with the transition operator.  In other words, $T_{\eta}$ converts Weyl quantization to an augmented quantization. 

From  (\ref{starodot}, \ref{Tnoqp}), we find the (generalized) Husimi star product\footnote{\,We mention that the Husimi star product (with $\eta=\hbar$) and the Moyal star product can be unified into a family of one-parameter star products, the so-called $s$-ordered products \cite{Cahill1969a,Lizzi2014}. For $s=0$, the Moyal product is recovered, while $s=1$ yields the Husimi star product.} 
\begin{align}
*_{T_\eta}\ =:\ *\sub{\,\eta}\ =\ \exp\left[\frac{i\hbar}2\left( \Lpq\Rpp - \Lpp\Rpq \right) + \frac{\eta}2\left( \sigma^2\Lpq\Rpq + \Lpp\Rpp/\sigma^2 \right)  \right]\ =\ *\sub{0}\,\odot\sub{\, \eta}\ .
\end{align}
Notice that $\odot\sub{\,\eta}^t = \odot\sub{\,\eta}$, so that, from (\ref{MoyalB}),
\begin{align}
\cM\sub{\,\eta}\ =\ \frac{*\sub{\,\eta}\, -\, *\sub{\,\eta}^{\, t}}{i\hbar}\ =\ \left( \frac{*\sub{0}\, -\, *\sub{0}^{\, t}}{i\hbar} \right)\,\odot\sub{\, \eta}\ =\ \cM\sub{0}\, \odot\sub{\, \eta}\  . 
\label{MMzOeta}
\end{align}
Therefore the classical limit produces a multiplicative modification of the Poisson bi-differential operator:
\begin{align}
\lim_{\hbar\to 0}\, \cM\sub{\eta}\ =\ \lim_{\hbar\to 0}\, \frac{(*\sub{0} - *\sub{0}^t )}{i\hbar}\, \odot\sub{\eta}\ =\  \cP\, \exp\left[\frac{\eta}2\left( \sigma^2\Lpq\Rpq + \Lpp\Rpp/\sigma^2 \right) \right]\  ,
\label{ClassStarEta}
\end{align}
rather than an augmentation of the form (\ref{limMaugP}).

\subsection{Damped quantization} 

Consider the simple harmonic oscillator, with Hamiltonian 
\begin{align}
H\ =\ \frac{1}{2m}p^2\ +\ \frac1 2 m\omega^2 q^2\ . 
\label{Hsho}\end{align}
Replace the Poisson bi-differential operator $\cP$ with \cite{Dito2006} 
\begin{align}
\cP\sub{\gamma}\ :=\ \cP\ -\ 2\gamma m \Lpp\,\Rpp \ . 
\label{Pgamma}\end{align} 
The canonical equations of motion are changed to
\begin{align}
\dot q\ =\ q\cP\sub{\gamma}H\ =\ p/m\ \ , \qq\dot p\ =\ p\cP\sub{\gamma}H\ =\ -m\omega^2 q \,-\, 2\gamma p\ . 
\label{DHO}\end{align}
The equation of motion of a damped harmonic oscillator results: 
\begin{align}
\ddot q\ =\ -\
\omega^2 q\ -\ 2\gamma\dot q\ , 
\label{qDHO}\end{align}
with $\gamma$ as the damping parameter. 

The same replacement in the Moyal star product produces a damped star product $*\sub{\gamma}$:
\begin{align}
*\sub{\gamma} = \exp\left\{\frac{i\hbar}2 \cP\sub{\gamma}\right\}\ =\  *\sub{0}\, \exp\left\{-i\hbar\gamma m\, \Lpp\,\Rpp\right\}\ =:\ *\sub{0}\, \odot\sub{\gamma}\ . 
\label{StarG}\end{align}
The transition operator 
\begin{align}
T\sub{\gamma}\ =\ \exp\left\{ -\frac{i\hbar\gamma m}2 \, \partial\sub{p}^2  \right\}
\label{Tgamma}\end{align} 
reproduces the star product $*\sub{\gamma}=*_{T_\gamma}$ when used in  (\ref{starodot}, \ref{Tnoqp})  \cite{Dito2006}. $T\sub{\gamma}$ therefore describes an augmented quantization of the simple harmonic oscillator, with classical damping introduced.   

Notice, however, that the transition operator $T\sub{\gamma}$ is not real.  This causes significant problems when the time evolution of quasi-probability distributions and observables is  considered \cite{Belchev2009}.  With $T\sub{\gamma}\not=\overline{T\sub{\gamma}}$, the damped $*$-product is non-Hermitian as a result:
\begin{align}
*\sub{\gamma}^\dagger = \exp\left\{\frac{i\hbar}2 \cP\sub{\gamma}\right\}^\dagger\ =\   *\sub{0}\,\, \odot\sub{\gamma}^\dagger\ =\ *\sub{0}\, \exp\left\{+i\hbar\gamma m\, \Lpp\,\Rpp\right\}\ \not=\ *\sub{\gamma}\ . 
\label{stardnonH}\end{align}
The dynamics is governed by the Moyal bi-differential operator (see (\ref{HeisMoyalEqn}) above), but with $*\sub{\gamma}^\dagger\not= *\sub{\gamma}$, 
\begin{align}
\cM\sub{\gamma}\ =\ \frac{*\sub{\gamma}-*\sub{\gamma}^t}{i\hbar}\ \not=\ \overline{\cM\sub{\gamma}}\ .
\label{Mgnotreal}\end{align}
Therefore, the reality of an observable, such as $f$ in (\ref{HeisMoyalEqn}), is not preserved in evolution because
\begin{align}
f\cM_{\gamma}H\ =\ f\cM_0H+i\gamma\hbar\pq\pp f\ .
\end{align}

Furthermore, by (\ref{cMzsin}), 
\begin{align}
\lim_{\hbar\to 0}\, \cM\sub{\gamma}\ =\ \lim_{\hbar\to 0}\, \frac{(*\sub{0} - *\sub{0}^t )}{i\hbar}\, \odot\sub{\gamma}\ =\ \lim_{\hbar\to 0}\, \frac{\sin\left(\hbar\, \cP\sub{\, 0}\,/ 2 \right)}{\hbar/2}\, \text{e}^{-i\hbar\gamma m \Lpp\Rpp}\ =\ \cP\ \not=\ \cP\sub{\gamma}\ .
\label{Mgclass}\end{align} 
That is, the damping disappears in the classical limit.  A quantization of the classical damped harmonic oscillator is not described after all \cite{Belchev2009}! 

The hopeful substitution \cite{Belchev2009} 
\begin{align}
\cM\sub{\gamma}\ \rightarrow\ \frac{*\sub{\gamma} - \overline{*\sub{\gamma}}}{i\hbar}\ =\ \frac{\sin\left(\hbar\cP\sub{\gamma}/2\right)}{\hbar/2}\ =\ \frac{*\sub{\gamma} - *\sub{-\gamma}^t}{i\hbar}
\label{MgtoIg}\end{align}
would fix both problems. But then the dynamical equation would become
\begin{align}
\dot f\ \rightarrow\ \frac{f\,*\sub{\gamma}\, H\ -\ H\, *\sub{-\gamma}\, f}{i\hbar}
\label{WtIt}\end{align}  
with formal solution 
\begin{align}
f(q,p;t)\ \rightarrow\ \exp\left[  \frac{it}{\hbar} H *\sub{\gamma}  \right]\,f(q,p;0)\, 
\exp\left[   *\sub{-\gamma}\frac{it}{\hbar} H  \right] \ .
\label{egWemg}\end{align}
The last expression is problematic, however.  Although $*\sub{\gamma}$ is associative when $\gamma$ is fixed, we find   
\begin{align}
\left( a\,*\sub{\gamma}\, b \right)\, *\sub{-\gamma}\, c\ \not=\ a\, *\sub{\gamma}\, \left( b\, *\sub{-\gamma}\, c \right)
\label{gmgNoAssoc}\end{align}
for phase-space obervables $a,b$ and $c$ \cite{Belchev2009}.  This non-associativity result follows from  
\begin{align} 
*\sub{-\gamma}(1+2, 3)\, *\sub{\gamma}(1, 2)\ \not=\ *\sub{\gamma}(1, 2+3)\, *\sub{-\gamma}(2, 3)
\label{StarGnum}\end{align}
in the notation of (\ref{starzassoc}, \ref{PREstarzassoc}). In turn, by (\ref{StarG}), (\ref{StarGnum}) is a consequence of 
\begin{align}
\cP\sub{-\gamma}(1+2, 3)\ +\ \cP\sub{\gamma}(1, 2)\ \not=\ \cP\sub{\gamma}(1, 2+3)\ +\ \cP\sub{-\gamma}(2, 3)\ .  
\label{Pgnum}\end{align}

Can something similar work better? Consider a generalization of the damped transition operator $T\sub{\gamma}$ of (\ref{Tgamma}): 
\begin{align}
T\sub{\gamma, \eta}\ =\ \exp\left\{ -\eta\gamma  m\, \partial\sub{p}^2  \right\}\ .
\label{Tgeta}\end{align}
As in the generalized Husimi quantization, the parameter $\eta$ has dimensions of action. It replaces $i\hbar/2$, so that it does not vanish in the $\hbar\to 0$ limit. Using $\eta\in \R$, equation (\ref{Tnoqp}) yields a Hermitian star product:
\begin{align}
*\sub{\gamma, \eta}\ =\ *\sub{0}\,\odot\sub{\gamma, \eta}\ =\ *\sub{0}\,\exp\big( -2\eta\gamma m\Lpp\Rpp  \big)\ =\ *\sub{\gamma, \eta}^\dagger\ \ \ (\eta\in \R)\ .
\label{StarGE}\end{align}
However, the classical limit still fails, as 
\begin{align}
\lim_{\hbar\to 0}\, \cM\sub{\gamma, \eta}\ =\ \lim_{\hbar\to 0}\, \frac{(*\sub{0} - *\sub{0}^t )}{i\hbar}\, \odot\sub{\gamma, \eta}\ =\  \cP\, \text{e}^{-2\eta\gamma m \Lpp\Rpp} \ .
\label{Mgeclass}\end{align} 
The damped Poisson bracket $\cP\sub{\gamma}$ of (\ref{Pgamma}) is not recovered. Notice this is true even if small $\gamma$ is considered,  
\begin{align}
\cP\, e^{-2\eta\gamma m \Lpp\Rpp} \ =\  \cP\,\Big[ 1 - 2\eta\gamma m\Lpp\Rpp + O(\gamma^2) \Big]\ \ .
\label{MgeclassSmall}\end{align} 
While $\cP\sub{\gamma}$ differs additively from $\cP$, equation (\ref{Mgeclass}) describes instead a {\it multiplicative} modification of $\cP$. A multiplicative modification of the classical limit was also found when using the Husimi transition differential operator in the previous subsection.

To progress, we need to understand what is and isn't possible in augmented quantization.  For that reason, we discuss the possibilities described by an arbitrary transition differential operator $T=T(\partial_q, \partial_p)$ in the next section.

\section{Augmented star products: generalities}
\label{sec: Augmented star products}

In this section, we consider modified star products in more general terms. The goal is to see if what has been learned from the examples leads us to results that pertain to any possible Hermitian, augmented star product. 

\subsection{Transition differential operators $T=T(\partial_q, \partial_p)$} 
\hfb
First, consider star products $*\sub{T}$ modified by a transition differential operator $T=T(\partial_q, \partial_p)$. $T$ describes the relation between $*\sub{T}$ and the reference Moyal star product $*_0$, see (\ref{starodot}, \ref{Tnoqp}).  We will restrict to real $T$, so that Hermitian star products $*\sub{T}$ result. 

In the classical limit $\hbar\to 0$, the dependence on $\hbar$ is crucial.  Consider an arbitrary transition differential operator  $T$; if a non-singular classical limit is to be found, we can write 
\begin{align}
T\ =\ T_0\ +\ \hbar\, T_1\ +\ \hbar^2\, T_2\ +\ \ldots\, ,\ \ {\text{ with}}\ \ \frac{\partial T_i}{\partial\hbar}\, =\, 0\ ,
\label{Thbarseries}
\end{align}
with no negative powers of $\hbar$.  As a consequence,\footnote{\,Notice that (\ref{Ohbar}) implies that the classical limit of the star product is $\lim_{\hbar\to 0} *\sub{T} = \odot\sub{T_0}$, as in (\ref{limstarnotid}).}
\begin{align}
*\sub{\, T}\ =\ *\sub{0}\, \odot\sub{\, T}\ =\ *\sub{0}\, \odot\sub{\, T_0}\ +\ \cO(\hbar^1)\ ,
\label{Ohbar}
\end{align}
with 
\begin{align}
\odot\sub{\, T_0}\ =\ T_0^{-1}(\Lpq,\Lpp)\, T_0(\Lpq+\Rpq, \Lpp+\Rpp)\, T_0^{-1}(\Rpq, \Rpp)\ .
\label{Tznoqp}\end{align}
Another critical observation is that 
\begin{align}\odot\sub{\, T}^{\, t}\ =\ \odot\sub{\, T}\ , 
\label{odotSymm}
\end{align}
by (\ref{Tnoqp}). Similarly, $\odot\sub{\, T_0}$ is symmetric, by (\ref{Tznoqp}).

The classical limit therefore yields
\begin{align}
\lim_{\hbar\to 0}\, \cM\sub{\, T}\ =\ \lim_{\hbar\to 0}\, \frac{(*\sub{\, T} - *\sub{\, T}^t )}{i\hbar}\ =\  \left(\lim_{\hbar\to 0}\, \frac{\left(*\sub{0} - *\sub{0}^t \right)}{i\hbar}\, \odot\sub{\, T} \right)\ =\ \left(\lim_{\hbar\to 0}\, \frac{\left(*\sub{0} - *\sub{0}^t \right)}{i\hbar} \right)\, \odot\sub{\, T_0} \  ,
\label{ClassMTnoh}
\end{align}
so that  
\begin{align}
\lim_{\hbar\to 0}\, \cM\sub{\, T}\ =\ \cP\,\odot\sub{\, T_0}\  .
\label{ClassMTPodot}
\end{align}

The classical limit is indeed modified, by that part of the transition differential operator that is independent of $\hbar$.  However, an augmentation (\ref{limMaugP}) of the desired additive form does not result - instead we find a multiplicative modification, (\ref{ClassMTPodot}). 

This is the kind of multiplicative modification described by the generalized Husimi product in (\ref{ClassStarEta}).  This is the only kind of modification that a real transition differential operator can describe. 

Of course, if we can write $\odot\sub{T_0}\approx 1 + \delta\odot\sub{T_0}$, we recover an approximate additive augmentation of the form  (\ref{limMaugP}), with $\delta\cP\approx \cP\,\delta\odot\sub{T_0}$.  But this is a very restrictive form.  For example, Hamilton's equations of motion for $q$ and $p$ can only be modified by the extra terms $q\cP\,\delta\hskip-2pt\odot\sub{T} H = \delta\hskip-2pt\odot\sub{T} \partial_pH$ and $p\cP\, \delta\hskip-2pt\odot\sub{T_0} H = -\delta\hskip-2pt\odot\sub{T_0} \partial_qH$. They vanish unless $\delta\odot\sub{T}$ contains terms  of the form $\Rpq^{\, m}$ or $\Rpp^{\, n}$, for some integer powers $m,n\in\N$. It never does, however, since  
\begin{align}
T_0\, \approx\, 1 + \theta_0\ \ \ \Rightarrow\ \ \ \delta\,\odot\sub{T_0}\, \approx\, -\,\theta_0(\Lpq, \Lpp)\, +\, \theta_0(\Lpq+\Rpq, \Lpp+\Rpp)\, -\,  \theta_0(\Rpq, \Rpp)\ ,
\label{dOdottheta}
\end{align}
by (\ref{Tznoqp}). We will therefore not consider this possibility further.  

Recall that  the Dito-Turrubiates transition operator (\ref{Tgamma})  resulted in a classical limit (\ref{Mgclass}) with no augmentation. 
To understand this, consider the transition differential operators on which (\ref{Tgamma}) was modelled: those that describe operator orderings that differ from the Weyl ordering, such as (\ref{BJTstar}).  They are necessarily $\hbar$-dependent, since different orderings can be related by the Heisenberg commutation relations. For them, however, a non-augmented classical limit is necessary if they are only to describe different quantizations of the same classical system. In other words, when these transition operators are applied, additional physical effects should not appear in the classical limit. 

We see now that the feature that the Dito-Turrubiates transition operator (\ref{Tgamma}) shares with transition differential operators describing changes in operator-ordering rules is 
\begin{align}
T_0\ =\ 1\ \ \ \Rightarrow\ \ \ \odot\sub{T_0}\ =\ 1\ .
\label{Tzone}
\end{align}

To summarize, only the $\hbar$-independent part $T_0$ of a transition differential operator $T(\partial_q, \partial_p)$ is relevant to the classical limit of the Moyal bi-differential operator $\cM\sub{T}$. If $T_0=1$, there is no augmentation, since the classical limit is unchanged by the transition operator. If $T_0\not=1$, however, there is an augmentation produced, but the classical Poisson bi-differential operator is not changed in the desired additive way, as in (\ref{limMaugP}). A multiplicative modification, 
(\ref{ClassMTPodot}), instead results. 

\subsection{Local transition operators $T=T(q,p,\partial_q, \partial_p)$}  
\hfb 
Since transition differential operators do not yield augmented star products, we will now consider generalized transition operators that depend on the phase-space coordinates: $T=T(q, p; \pq, \pp)$.  

As already noted, this is a radical step, but there are precedents. Generalizations in a similar spirit were previously discussed in \cite{Crehan1989, Pinzul2008}.\footnote{Although ref.~[14] does not  make explicit use of transition operators or star products, it does work in a mathematically equivalent formalism.  It should be noted, however, that only (``local'') quantizations dependent on $\hbar$ were considered. Ref. [15] focused on using transition operators to gauge the star product by expanding in powers of $\hbar$.}  

\subsubsection{Star product}
\hfb 
In the more general case, we conjecture that (\ref{starTFG}) is solved by 
\begin{align}
*\sub{\, T}(1,2)\ =\ T(1,2)\, *\sub{\, 0}\hskip-2pt(1,2)\,\, T^{-1}(1)\, T^{-1}(2)\ ,
\label{starTTstar}\end{align}
where now 
\begin{align}
T(1,2)\ :=\ T\Big( (q_1+q_2)/2, (p_1+p_2)/2; \partial_{q_1}+\partial_{q_2}, \partial_{p_1}+\partial_{p_2} \Big)\, ,\  {\text{when}}\  T=T(q, p; \pq, \pp)\ .
\label{Tqp}\end{align}
The bi-differential operator $*\sub{\, T}$ is obtained from $*\sub{\, T}(1,2)$ by identifying $(q_1, p_1) = (q_2, p_2) = (q, p)$: 
\begin{align}
*\sub{\, T}\,  =\, \I(1,2)\,\, *\sub{\, T}(1,2)\ \ . 
\end{align} 
Notice that when $T=T(\partial_q, \partial_p)$ is a differential operator, the simpler result (\ref{Tnoqp}) is recovered. An explicit, general expression for $*\sub{\, T}$, however,  in terms of $q, p, \Lpq, \Rpq, \Lpp$ and $\Rpp$, is out of reach. 

The result  (\ref{starTTstar}, \ref{Tqp}) appears to be new.  The formulas can be given some justification as follows. We must show that 
\begin{align}
\I(1,2)\, T(1,2)\, K(1,2)\ =\ T\, \I(1,2)\, K(1,2)\ ,
\end{align}
for arbitrary $K(1,2):= K(q_1,p_1; q_2, p_2)$.  

Using the commutation relations 
\begin{align}
[\partial_q, q]\ =\ [\partial_p, p]\ =\ 1\ ,
\end{align}
we can rewrite the transition operator in the form 
\begin{align}
T\ =\ T(q,p,\partial_q, \partial_p)\ =\ \sum\, t_{m,n}(q,p)\, \partial_q^{\, m}\, \partial_p^{\, n}\ . 
\end{align}
It follows that $T(1,2)$ obeys
\begin{align}
T(1,2) \ =\ \sum\, t_{m,n}\left(\frac{q_1+q_2}{2}, \frac{p_1+p_2}{2}\right)\, (\partial_{q_1}+\partial_{q_2})^{\, m}\,\, (\partial_{p_1}+\partial_{p_2})^{\, n}\ ,
\end{align}
since 
\begin{align}
[\, \partial_{q_1}+\partial_{q_2}, (q_1+q_2)/2\, ]\ &=\ [ \partial_q, q]\ =\ 1\ \ \&\nn 
[\, \partial_{p_1}+\partial_{p_2}, (p_1+p_2)/2\, ]\ &=\ [ \partial_p, p]\ =\ 1\ . 
\end{align} 
A similar result holds if $(q_1+q_2)/2$ and $(p_1+p_2)/2$ are replaced by $\alpha q_1 + (1-\alpha)q_2$ and $\beta p_1 + (1-\beta) p_2$, for any $0\leq \alpha, \beta \leq 1$, since the necessary commutation  relations are obeyed. We choose the most symmetrical solution here, however, for simplicity. 

Since $\I(1,2) t_{m,n}\big( (q_1+q_2)/2, (p_1+p_2)/2 \big) = t_{m,n}(q,p)$, what remains to be shown is 
\begin{align}
\I(1,2)\, (\partial_{q_1}+\partial_{q_2})^{\, m}\,\, (\partial_{p_1}+\partial_{p_2})^{\, n}\, K(1,2)\ =\ \partial_q^{\, m}\, \partial_p^{\, n}\, \I(1,2)\, K(1,2)\ .
\end{align} 
Assuming we can expand 
\begin{align}
K(1,2)\ =\ K(q_1, p_1; q_2, p_2)\ =\ \sum\, k_{a_1,b_1; a_2,b_2}\, q_1^{a_1} p_1^{b_1} q_2^{a_2} p_2^{b_2} \ , 
\end{align}
we need only show that 
\begin{align}
\I(1,2)\, (\partial_{q_1}+\partial_{q_2})^{\, m}\,\, (\partial_{p_1}+\partial_{p_2})^{\, n}\,\, q_1^{a_1} p_1^{b_1} q_2^{a_2} p_2^{b_2}\ =\ \partial_q^{\, m}\, \partial_p^{\, n}\, q^{a_1+a_2}\, p^{b_1+b_2}\ ,  
\end{align}
which reduces to showing 
\begin{align}
\I(1,2)\, (\partial_{q_1}+\partial_{q_2})^{\, m}\,\, q_1^{a_1}  q_2^{a_2}\ =\ \partial_q^{\,\, m}\, q^{a_1+a_2}\ .   
\label{reduced}\end{align} 
But this last equation is satisfied. 

It is important to note that the associativity of $*\sub{\, T}$ follows from that of the Moyal star product $*\sub{\, 0}$, for any invertible $T$, whether it depends on $q$ and $p$ or not. One obtains 
\begin{align}
\Big( Tf\, *\sub{\, T}\, Tg \Big)\, *\sub{\, T}\, Th\ =\ Tf\, *\sub{\, T}\, \Big( Tg \, *\sub{\, T}\, Th \Big)\ , 
\label{starTassoc}\end{align}
by applying (\ref{starT}) twice to (\ref{starzassoc}). 

\subsubsection{Augmented equations of motion}
\label{sec: Augmented equations of motion}
\hfb 
In this section, we will use  local transition operators $T=T(q,p; \partial_q, \partial_p)$ to try to find star products that yield classical limits augmented by additional physics. We will focus on the equations of motion for the phase-space coordinates $q, p$. 

Let $x$ denote either $q$ or $p$. Precisely, we will ask that
\begin{align}
\dot x\ =\ \lim_{\hbar\to 0}\, \{ x, H \}\sub{*_T}\  
\label{LimEqMx}
\end{align}
describes the augmented quantization in the classical limit.

With $T=T(q,p; \partial_q, \partial_p )$ a local transition operator, we have no general formula for $*\sub{\, T}$ written directly in terms of left- and right-derivatives. It is therefore easiest to work with $T$ directly and use (\ref{starTFG}) to rewrite (\ref{LimEqMx}) as 
\begin{align}
\dot x\ =\ \lim_{\hbar\to 0}\, T\Big(\,  \{ T^{-1}x, T^{-1}H \}\sub{\, *_0} \,\Big)\ =\   
T\Big(\,  \{ T^{-1}x, T^{-1}H \}\,\Big)\ ,
\label{LimEqMxT}
\end{align}
where we have used $\partial T/\partial\hbar = 0$ and (\ref{ClimPB}). 

We will consider augmentations that are weak, by writing 
\begin{align}
T\ =\  \approx\ 1\,+\, \theta\ ,\ \ \ \ T^{-1}\ \approx\ 1 \,-\, \theta\ , 
\label{weakTtheta}
\end{align}
with $\theta$ a bi-differential operator. Then (\ref{LimEqMxT}) becomes 
\begin{align}
\dot x\ -\ \{ x, H \}\ \approx\ \theta\big(\, \{ x, H \} \,\big)  \ -\ \{\theta(x), H\}\ -\ \{ x, \theta(H) \}\ , 
\label{ThetaAug}
\end{align}
where the augmenting terms are all on the right-hand side.  It is helpful, perhaps, to rewrite them in the notation of (\ref{shorth}).  Denoting the  terms augmenting the equations of motion by $\cA\sub{\theta}(x)$, we have
\begin{align}
\cA\sub{\theta}(x)\ =\ \I(1,2)\,\Big\{  \theta(1+2)\,\cP(1,2) \, -\, \cP(1,2)\, \big[ \theta(1) + \theta(2)   \big] \Big\}\, x(1)\,H(2)\ .
\label{AugShort}
\end{align}

From these expressions, one sees that $\theta = \beta \partial_{x'}$ ($\beta$ a constant, and $x'=q$ or $p$) produces no augmentation.  Furthermore, no term contributes that is higher order in derivatives $\partial_q, \partial_p$, if it is multiplied by a constant. A multiplicative function of $q$ and $p$ in $\theta$ is necessary - this confirms that a local transition operator is required. 

Consider then the ansatz 
\begin{align}
\theta\ =\ \sum_{{m,n}\atop{1\leq m+n}}\, \theta\sub{m,n}\, \partial_{q}^{\, m}\, \partial_{p}^{\, n}\ , 
\label{ThetaQuad}
\end{align}
producing 
\begin{align}
\cA\sub{\theta}(q)\ =&\  -\{\theta\sub{1,0}, H\}\ -\sum_{m,n} \left(\partial_p\theta\sub{m,n}\right) \left( \partial_q^{\, m} \partial_p^{\, n}H\right)\ =\  -\{\theta\sub{1,0}, H\}\ -\ \big[\, (\partial_p\theta) H \,\big] , \nn 
\cA\sub{\theta}(p)\ =&\  -\{\theta\sub{0,1}, H\}\ +\sum_{m,n} \left(\partial_q\theta\sub{m,n}\right) \left( \partial_q^{\, m} \partial_p^{\, n}H\right)\ =\  -\{\theta\sub{0,1}, H\}\ -\ \big[\, (\partial_q\theta) H \,\big] \ .
\label{AugThetaqp}
\end{align}
\subsubsection{Example: weakly damped harmonic oscillator}
\hfb
As an example, consider augmenting the simple harmonic oscillator by introducing weak damping, with damping coefficient $\gamma$.  To produce the equations of motion (\ref{DHO}) for the damped oscillator, we need 
\begin{align}
\cA\sub\theta(q)\ =\ 0\ ,\ \ \ \cA\sub\theta(p)\ =\ -2\gamma p\ ,
\label{DHOaug}
\end{align} 
when the Hamiltonian (\ref{Hsho}) is used in (\ref{AugThetaqp}). 
We find that 
\begin{align}
\theta\ =&\ \theta_{0,1}\,\partial_p\ +\ \theta_{0,2}\,\partial_p^{\, 2}\ , \nn
\theta_{0,1}\ =&\ 2\frac\gamma\omega\, \int \arctan \left( \frac{m\omega q}p \right)\, dp\ ,\nn
\theta_{0,2}\ =&\ -\frac\gamma\omega\, \left\{ 2mH\,\arctan \left( \frac{m\omega q}p \right)\ +\ m\omega q p \right\}\ , 
\label{TgaLoc}
\end{align}
yields (\ref{DHOaug}). Here $\int\arctan({m\omega q}/p)\, dp$ indicates the indefinite integral.\footnote{${}\,$For simplicity, we have not written the somewhat more general solution we obtained for  $\theta$ of (\ref{AugThetaqp}, \ref{DHOaug}). It is recorded in Appendix C of \cite{Robbins2017}. Also written there is a $\theta$ producing quadratic damping.} 

This example demonstrates that local transition operators may be capable of  producing star products describing systems with additional classical effects. A Hermitian star product incorporating weak damping, at least in the equations of motion for phase-space coordinates $q$ and $p$, is described by the local transition operator of (\ref{weakTtheta}, \ref{TgaLoc}). 

However, we have not worked out the modified Moyal bi-differential operator $\cM$  (\,see (\ref{MoyalB}, \ref{cMstar})\,) in general, and we have no explicit formula for the star product in terms of left- and right-derivatives. The result (\ref{TgaLoc}) is quite complicated, while only producing the desired $q$ and $p$ equations of motion, and being only valid for weak coupling. Furthermore, the damping it describes applies only to the simple harmonic oscillator - a different form would be required for the damping of a different system.  

All this suggests that even if local transition operators may produce quantizations augmented by classical effects, their use may turn out to be too unweildy to be practical in what is an effective description of certain physical systems.

\section{Conclusion} 

Our goal was to investigate the possibility, suggested in \cite{Dito2006}, of introducing additional classical physics during phase-space quantization. To do this, we analyzed transition operators describing Hermitian quantizations related to the Weyl-Wigner quantization, with its Moyal star product. Can transition operators yield such augmented quantizations? 

By examining the classical limit, we showed that transition differential operators $T(\pq, \pp)$ cannot. Only the $\hbar$-independent part $T_0$ of the transition operator is relevant for the classical limit. If $T_0=1$, then there is no augmentation: the classical limit recovers the usual Poisson brackets, with no extra classical physics. If $T_0\not=1$, the classical physics is indeed modified, but in a multiplicative, rather than additive way. 

So, augmented quantization using  transition differential operators $T=T(\pq, \pp)$ does not produce a classical limit of the desired form (\ref{limMaugP}).

In retrospect, we see that the Dito-Turrubiates \cite{Dito2006, Belchev2009} transition operator (\ref{Tgeta}) is an example for which $T_0=1$.  Also, the (generalized) Husimi \cite{Husimi1940, Takahashi1989} transition operator  (\ref{Teta}) illustrates the other case, with $T_0\not=1$, but describing a multiplicative modification of the classical physics. 

For a more thorough treatment, we also considered a significant generalization. Local transition operators $T=T(q, p, \pq, \pp)$, dependent on the phase-space point, were also examined. We managed to construct a single example, a real (local) transition operator (\ref{weakTtheta}, \ref{TgaLoc}) for a Hermitian star product that introduces a weak damping into the $q, p$ equations of motion of the simple harmonic oscillator. 

However, the analysis is significantly more difficult when using local transition operators. Only the equations of motion for the phase-space coordinates were examined; the modified Moyal bi-differential operator was not worked out.  Furthermore, the transition operator (\ref{weakTtheta}, \ref{TgaLoc}) is quite complicated.  At some point, an effective description is not practical if it is too involved.  

The weak-coupling result we found is also of limited validity, and the transition operator has the undesirable feature of being specific to the harmonic oscillator. It would have to take a different form to introduce damping into a different quantum system.  This contrasts sharply with  the standard procedure: to incorporate an additional physical effect into different systems, an identical term is added to the different Hamiltonians.  

We conclude, therefore, that the significant generalization to local transition operators yields a rather unwieldy machinery.  In its original form, augmented quantization with a Hermitian star product does not work, and with local transition operators, it is not clear that it is practical.

\vskip1cm
\noindent{\bf Acknowledgements}\hfb 
This research was supported in part by a Postgraduate Scholarship (MR) held at the University of Lethbridge and a Discovery Grant (MW, grant number RGPIN 5809/2015) from the Natural Sciences and Engineering Research Council (NSERC) of Canada. Some of the results of this paper were already reported in \cite{Robbins2017}, available at https://www.uleth.ca/dspace/handle/10133/4923.

\vfill\eject



\bibliographystyle{unsrt}
\bibliography{RobbinsWalton2018References}



\end{document}